\documentclass[twocolumn,aps,apl,amsmath,amssymb]{revtex4-1}
\usepackage{graphicx, dcolumn, amsmath}
\begin{document}
 \preprint{}

\title{Band gap control via tuning of inversion degree in CdIn$_2$S$_4$ spinel}

\author{Yohanna Semin\'{o}vski$^{1,2}$}
\email{seminovski@etsit.upm.es}
\author{Pablo Palacios$^{1,3}$}
\author{Perla Wahn\'{o}n$^{1,2}$}
\author{Ricardo Grau-Crespo$^{4}$}
\affiliation{$^{1}$ Instituto de Energ\'{i}a Solar, Universidad Polit\'{e}cnica de Madrid UPM, Ciudad Universitaria, 28040, Madrid, Spain}
\affiliation{$^{2}$ Dpt. TEAT, ETSI Telecomunicaci\'{o}n, UPM, Ciudad Universitaria. 28040, Madrid, Spain}
\affiliation{$^{3}$ Dpt. FyQATA, EIAE, UPM, Pz. Cardenal Cisneros 3, 28040, Madrid, Spain}
\affiliation{$^{4}$ Department of Chemistry, University College London, 20 Gordon Street, London WC1H 0AJ, UK}
\date{\today}

\begin{abstract}
Based on theoretical arguments we propose a possible route for
controlling the band-gap in the promising photovoltaic material
CdIn$_2$S$_4$. Our \textit{ab initio} calculations show that the experimental degree
of inversion in this spinel (fraction of tetrahedral sites occupied by
In) corresponds approximately to the equilibrium value given by the
minimum of the theoretical inversion free energy at a typical
synthesis temperature. Modification of this temperature, or of the cooling rate after synthesis,
is then expected to change the inversion degree, which in turn sensitively tunes the electronic
band-gap of the solid, as shown here by accurate screened hybrid functional calculations.
 \end{abstract}

\maketitle

 The cadmium/indium thiospinel CdIn$_2$S$_4$ is a photosensitive semiconductor with excellent light absorption properties in the visible range of the spectrum, and has attracted considerable attention in recent years due to its potential  applications in photocatalysis, high-efficiency solar cells, light-emitting diodes, and optoelectronic devices \cite{kale06, apte10, aguilera10, lucero11}. A detailed understanding of the factors controlling its electronic band structure is required in order to optimize these applications. The band gap of pure CdIn$_2$S$_4$ is indirect and values between 2.1 eV and 2.4 eV (between 2.5 and 2.7 eV for the direct gap) have been reported by different authors \cite{lb}. The electronic and optical properties of the system at room temperature seem to depend on whether the crystal is annealed or quenched from the synthesis temperature, which is possibly related to changes in the cation distribution \cite{kulikova88}.

In a \lq normal' spinel, the 2+ cations are located in the tetrahedral sites and the 3+ cations occupy octahedral sites, while deviation from this distribution is called \lq inversion'. For CdIn$_2$S$_4$ it is difficult to refine the Cd/In occupancies of the tetrahedral and octahedral sites from standard diffraction measurements because Cd$^{2+}$ and In$^{3+}$ are isoelectronic, but Raman experiments suggest that some level of inversion is present \cite{shimizu75, ursaki02}. We can write the formula unit as (Cd$_{1-x}$In$_x$)[In$_{2-x}$Cd$_{x}$]S$_4$, where ( ) represent the tetrahedral sites, [ ] the octahedral sites, and $x$ is the degree of inversion ($x$=0 is the normal spinel and $x$=1 is the fully inverse spinel). The experimental value of $x$ is 0.20 \cite{ursaki02}.

In this Letter we present theoretical results showing that (\emph{i}) the experimental degree of inversion degree roughly corresponds to thermodynamic equilibrium at the formation temperatures of the thiospinel, thus suggesting that the distribution of cations could be modified by changing the synthesis temperature, and (\emph{ii}) that even a moderate change in the degree of inversion leads to significant change in the band gap of the material, which indicates a promising route for tuning the light absorption properties of the semiconductor.

 We have performed density functional theory (DFT) calculations of CdIn$_2$S$_4$ with different inversion degrees, using the Vienna Ab Initio Simulation Program (\texttt{VASP}) \cite{kresse99}. We employed both the Perdew-Burke-Ernzerhof (PBE) exchange-correlation functional \cite{pbe96} and its modified version for solids (PBEsol) \cite{perdew08}, which  improves the description of cell geometries and of phonon frequencies in solids, compared to the standard PBE functional \cite{pierre11}. For the direct CdIn$_2$S$_4$, PBEsol gives an equilibrium cell parameter ($a$=10.860 \AA) that is closer than the PBE result (a=11.020 \AA) to the experimental value ($a_{exp}$=10.831  \AA, which can be obtained by extrapolation to absolute zero of the temperature dependence measured by Kistaiah et al. \cite{kistaiah82}). The convergence of energies with respect to cutoff energies, reciprocal space sampling density and other precision parameters was checked carefully. The CdIn$_2$S$_4$ primitive cell contains two formula units, which allows us to consider inversion degrees $x$ = 0, 0.5 and 1. For each value of $x$ there is only one symmetrically different configuration of cations, so we calculate the inversion energy as $\Delta E_{conf} (x)=E(x)-E(0)$, where the subscript \emph{conf} indicates that this is the configurational contribution only (vibrational contributions are discussed below).

\begin{figure}[t]
\includegraphics[width=8.5cm]{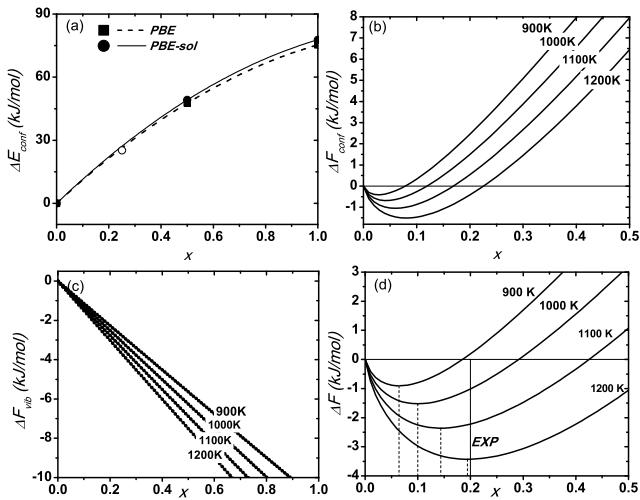}
\caption{\label{fig:uno} a) Inversion energies as obtained from PBE and PBEsol calculations; the open circle corresponds to the ensemble average in a double cell with $x$=0.25.
 b) Configurational and  c) vibrational contributions to the inversion free energy. d) Total inversion free energy; the vertical line marks the experimental value of the inversion degree.}
\end{figure}

 A quadratic dependence of the inversion energy with $x$ has been empirically recognised by Kriessman and Harrison \cite{kriessman56} and theoretically justified by O'Neill and Navrotsky \cite{oneill83} based on the linear dependences with $x$ of both the cell parameter $a$ and the anion position parameter $u$. Therefore, our three calculated inversion energies allow us to interpolate for any value of $x$ using a quadratic function, as shown in Fig \ref{fig:uno} a).  The PBE and PBEsol results are very similar in terms of energy, therefore the discussion below refers to the PBEsol results, unless otherwise stated.  The inversion energies are positive for the whole range of $x$, and the curvature is slightly negative. In order to test the approximation of using only one configuration for the calculation of the energy for each  $x>0$, we have calculated the inversion energy for $x$=0.25 by taking the symmetry-adapted ensemble average of all different cation configurations \cite{grau07} in a cell doubled along one axis. The result, represented as an empty circle in Fig. \ref{fig:uno} a), is in good agreement with our quadratic interpolation based on the primitive cell results.

We can now estimate the configurational free energy of inversion $\Delta F_{conf}=\Delta E_{conf}-T \Delta S_{conf}$, where:

\begin{eqnarray}
\label{eq:sconf}
S_{conf}=-R [xlnx+(1-x)ln(1-x)+ \nonumber\\
         xln\frac{x}{2}+(2-x)ln( 1-\frac{x}{2} ) ]
\end{eqnarray}

is the ideal configurational entropy of inversion \cite{navrotsky67, tielens06}. In practice, some excess (non-ideal) contributions to the configurational entropy can be expected, but our test calculations for the ensemble of configurations with $x$=0.25 in the double cell show that these excess contributions are indeed small: the difference between the temperature-dependent entropy (as calculated using Boltzmann statistics, e.g. \cite{benny09}) and the maximum entropy for the given cell and composition, is only 2\%.  In the absence of inversion energies, the minimum of the inversion free energy corresponds to the maximum of the configurational entropy, which occurs at $x=2/3$ (full disorder of the cations among \emph{all} sites).  However, the positive and relatively high inversion energies imply that the minima of the inversion free energy actually occur at much lower values of $x$ (e.g. $x$=0.04 at $T$=1000 K), as shown in Fig \ref{fig:uno} b). Therefore, according to the analysis so far, which ignores vibrational effects, the CdIn$_2$S$_4$ spinel would be expected to be almost fully direct, while experimental measurements suggest inversion degrees of  $\sim$20\%  \cite{ursaki02}.

 In order to discuss the effect of vibrations in the thermodynamics of inversion, we now examine the phonon modes in CdIn$_2$S$_4$ with different degrees of inversion. For these calculations, we employ density functional perturbation theory, as implemented in \texttt{VASP}. We first compare the calculated zone-centre frequencies for the direct spinel with the experimental values measured using infrared and Raman spectroscopy \cite{yamamoto73, shimizu75}.  TABLE \ref{tab:table4} shows that the vibrational modes are well described by our PBEsol calculations, with average discrepancy of only 2.5\%.

\begin{table}[!b]
\caption{\label{tab:table4}%
Calculated zone-centre phonon frequencies of normal CdIn$_2$S$_4$ spinel in comparison with experimental values \cite{yamamoto73, shimizu75} }
\begin{ruledtabular}
\begin{tabular}{ccddd}
Mode&$\tilde{\nu}_{exp}$$(cm^{-1})$&
\multicolumn{1}{c}{\textrm{$\tilde{\nu}_{theo}$}$(cm^{-1})$}\\
\hline
$T_{1u}$ (IR)           & 68&67\\
$T_{2g}$ (Raman)        & 93&92\\
$T_{1u}$ (IR)           &171&171\\
$E_g$    (Raman)        &185&189\\
$T_{1u}$ (IR)           &215&221\\
$T_{2g}$ (Raman)        &247&239\\
$T_{1u}$ (IR)           &307&296\\
$T_{2g}$ (Raman)        &312&297\\
$A_{1g}$ (Raman)        &366&355\\
\end{tabular}
\end{ruledtabular}
\end{table}

We then calculated the phonon frequencies in a $2\times2\times2$ supercell, which is equivalent to a $\Gamma$-centered $2\times2\times2$ sampling of the reciprocal space, thus allowing (linear) dispersion of the modes. From the resulting frequencies we obtained the vibrational contributions to the inversion free energy in the harmonic approximation:

  \begin{eqnarray}
  \Delta F_{vib}=k_BT \{ \sum ln \left[2sinh \left(\frac{h \nu_i(x)}{2k_BT}\right) \right]-  \nonumber\\
                         \sum ln \left[2sinh \left(\frac{h \nu_i(0)}{2k_BT}\right) \right]  \}
\end{eqnarray}

 for configurations with inversion degrees $x$=0, 0.5 and 1.  Fig. \ref{fig:uno} c) shows that the vibrational contribution to the free energy exhibits an almost linear variation with $x$. The negative slope means that vibrational effects will shift the equilibrium inversion degree towards values higher than those expected based on configurational contributions only. The total (configurational + vibrational) free energy is then plotted in Fig. \ref{fig:uno}d) as a function of $x$ for temperatures between 900 and 1200 K.  At a temperature $T$= 1200 K, which is typical for the synthesis of this type of material \cite{kistaiah82, gibart74}, the equilibrium degree of inversion is very close to the experimentally observed value $x$=0.20. This result suggests that the degree of inversion in CdIn$_2$S$_4$ is thermodynamically controlled during the solid formation.

 The sensitivity of the equilibrium inversion to the temperature indicates that it is possible to tune the cation distribution via temperature control during the sample preparation. For example, if the equilibration temperature is reduced from 1200 K to 1100 K, the degree of inversion decreases to $x$=0.15. From these results, it can also be expected that the degree of inversion will be higher when the sample is quenched (cooled rapidly) instead of annealed (cooled slowly) after its formation. In the latter case, the slow cooling will allow the equilibration of the cation distribution at lower temperatures. Experimental studies have indeed shown a variation in the electronic and optical properties of CdIn$_2$S$_4$ with the cooling rate after synthesis \cite{kulikova88}.

\begin{figure}[t]
\includegraphics[width=8cm]{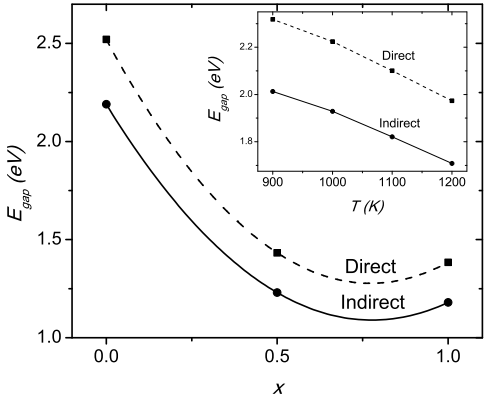}
\caption{\label{fig:dos} Direct and indirect band gaps as a function of inversion degree, from screened hybrid functional calculations. In the inset, the band gaps as a function of the configurational equilibration temperature.}
\end{figure}

We therefore consider the effect that a change of inversion degree has in the electronic properties of the material. The electronic structure was calculated here using the Heyd-Scuseria-Ernzerhof  (HSE06) screened hybrid  functional, which includes 25\% of Hartree-Fock exchange \cite{hse06}, yielding better band gap predictions (although at a higher computational cost) than standard DFT functionals \cite{henderson}. In fact, recent work has  shown that HSE06 gives band structures for Mg/In and Cd/In thiospinels that are very similar to those obtained from more computationally demanding many-body techniques \cite{lucero11}. The calculated values for the direct and indirect band gap are close to, although somewhat lower than, the experimental values measured in samples with degree of inversion $x$=0.20 (2.1-2.4 for indirect and 2.5-2.7 for direct band gap \cite{lb}).  The effect of inversion on the band gap is found to be quite drastic, with a decrease of almost 1 eV in both the direct and indirect band gaps when $x$ changes from 0 to 0.5 (Fig. \ref{fig:dos}).

This result means that any small change in the inversion degree, achieved via temperature control during the sample preparation, will be reflected significantly in the electronic and optical properties of the material. The connection between the band gap and the equilibration temperature for the cation distribution is illustrated in the inset of Fig. \ref{fig:dos}. Lower temperatures of formation, or slower cooling rates after sample preparation, should lead to wider band gaps. Therefore our simulations results indicate a possible route to tune the electronic properties of this interestic photovoltaic material.

The Spanish effort was supported by the Ministerio de Ciencia y Innovacion: Consolider Ingenio2010 GENESIS-FV (Grant Nº CSD2006-04), FOTOMAT (Grant No. MAT2009-14625-C03-01), and the Comunidad de Madrid NUMANCIA2 project (S-2009ENE-1477).
We thank the UK's  Materials Chemistry Consortium  for allowing access to the HECToR supercomputing service via EPSRC grant EP/F067496.

\end{document}